\documentclass[10pt]{IEEEtran}
\usepackage[noadjust]{cite}
\usepackage{amsmath}
\usepackage{amssymb}
\usepackage{mathtools}
\usepackage{bbm}
\usepackage{bm}
\usepackage{comment}
\usepackage{tabularx,booktabs}
\usepackage{caption}
\usepackage{subcaption}
\usepackage[capitalize]{cleveref}
\usepackage{pgfplots}
\pgfplotsset{compat=1.15}
\usepgfplotslibrary{fillbetween}
\usetikzlibrary{patterns,arrows,plotmarks}
\usepgfplotslibrary{groupplots}
\pgfdeclarelayer{background}
\pgfsetlayers{background,main}
\usetikzlibrary{automata,positioning}
\usetikzlibrary{decorations}
\usetikzlibrary{decorations.markings}
\makeatletter
\tikzset{
  nomorepostactions/.code={\let\tikz@postactions=\pgfutil@empty},
  mymark/.style 2 args={decoration={markings,
    mark= between positions 0 and 1 step 0.1 with{ 
      \tikzset{#2,every mark}\tikz@options
        \pgftransformresetnontranslations
        \pgfuseplotmark{#1}%
      },  
    },
    postaction={decorate},
    /pgfplots/legend image post style={
        mark=#1,mark options={#2},every path/.append style={nomorepostactions}
    },
  },
}
\makeatother

\definecolor{color0}{RGB}{142, 202, 230}
\definecolor{color1}{RGB}{2, 48, 71}
\definecolor{color2}{RGB}{251, 133, 0}
\usepackage{algorithm}
\usepackage[noend]{algpseudocode}

\begin{document}
\title{Prediction of mmWave/THz Link Blockages through Meta-Learning and Recurrent Neural Networks}

\author{Anders E. Kal{\o}r,~\IEEEmembership{Student Member,~IEEE,} Osvaldo Simeone,~\IEEEmembership{Fellow,~IEEE,} and Petar Popovski,~\IEEEmembership{Fellow,~IEEE}%
\thanks{This work has been in part supported by the Danish Council for Independent Research, Grant Nr. 8022-00284B (SEMIOTIC).}
\thanks{A. E. Kal{\o}r and P. Popovski are with the Department of Electronic Systems, Aalborg University, Denmark (email: \{aek,petarp\}@es.aau.dk).}
\thanks{O. Simeone is with the Centre for Telecommunications Research, Department of Engineering, King’s College London, United Kingdom (email: osvaldo.simeone@kcl.ac.uk).}}

\maketitle

\begin{abstract}
Wireless applications that rely on links that offer high reliability depend critically on the capability of the system to predict link quality within a given time interval. This dependence is especially acute at the high carrier frequencies used by mmWave and THz systems, where the links are susceptible to blockages. Predicting blockages with high reliability requires a large number of data samples to train effective machine learning modules. With the aim of mitigating data requirements, we introduce a framework based on \emph{meta-learning}, whereby data from distinct deployments are leveraged to optimize a shared initialization that decreases the data set size necessary for any new deployment. Predictors of two different events are studied: (1) at least one blockage occurs in a time window, and (2) the link is blocked for the entire time window. The results show that an RNN-based predictor trained using meta-learning is able to predict blockages after observing fewer samples than predictors trained using standard methods.
\end{abstract}
 
\begin{IEEEkeywords}
Blockage prediction, mmWave communication, meta-learning.
\end{IEEEkeywords}

\section{Introduction}
Due to its extreme bandwidth and delivery of high rates, highly directional millimeter wave (mmWave) and THz communication links are attractive options for many wireless applications including virtual reality (VR)~\cite{elbamby2018toward} and Industrial Internet-of-Things (IIoT)~\cite{sachs2019boosting}. However, the susceptibility of directional links to blockages makes it challenging to deploy the technology for time-sensitive applications with high reliability requirements. For instance, many IIoT applications, such as motion control systems, rely on frequent periodic transmissions of short packets that must be delivered with low latency (sub-millisecond to few milliseconds) and high reliability (up to $1-10^{-8}$)~\cite{3gpp22104_17_4_0}. Moreover, the system must be down only for a short duration known as the \emph{survival time}. The impact of blockages needs to be reduced by implementing reactive or preventive mechanisms in the communication system, such as searching for alternative communication paths or adopting a more robust control strategy to avoiding unnecessary system shutdowns. While reactive mechanisms bring a detection delay, preventive measures, such as beam training, incur high overhead as they need to be repeated at reoccurring intervals.

To achieve both low latency and low overhead in applications such as VR and IIoT, recent studies have considered the problem of \emph{blockage prediction} with the aim of predicting when blockages will occur, and to proactively initiate countermeasures. Due to the lack of mathematical models for blockages, the predictors are often based on machine learning models, which use metrics from the communication link as well as external features such as visual and location information~\cite{she2021urllc}. A range of works have applied neural networks to the problem of using sub-6 GHz channels to predict blockages in the mmWave bands~\cite{burghal2019deep,alrabeiah20,ali19}. The use of Recurrent Neural Networks (RNNs) to predict beam blockages in scenarios with user mobility has been studied in~\cite{alkhateeb18,shah20,hussain20,liu20}, which learn the spatial and/or temporal correlation of blockages. Furthermore, visual features has been shown to increase the accuracy of such predictors~\cite{alrabeiah20,charan21}. Other works predict blockages using diffraction effects, etc.~\cite{wu21}, and \cite{samarakoon20} uses survival analysis to predict the probability of experiencing a blockage based on a recent observation window.

\begin{figure}
  \centering
  \includegraphics{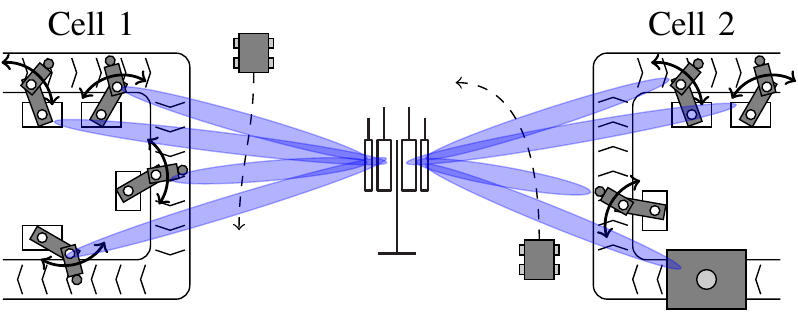}
  \caption{An example scenario comprising two manufacturing cells with robots that are connected to a base station through mmWave links. The mmWave links are subject to blockages from the robotic arms, as well as by moving vehicles.}
  \label{fig:introfig}
\end{figure}

One of the main challenges associated with machine learning predictors is that they rely on large amounts of training samples from their specific deployment in order to accurately learn the dynamics of the blockages. In this paper, we address this problem using \emph{meta-learning}, which has previously been applied successfully to wireless systems~\cite{simeone20}. Meta-learning allows one to exploit observations from a set of deployments to optimize the \emph{inductive bias}, allowing predictors, such as a neural networks, to be trained for a new deployment using much fewer samples than would otherwise be required.

As an example, consider an indoor industrial scenario as depicted in \cref{fig:introfig}, where blockages are predominantly due to the movement of agents such as robots and conveyor belts. Although the blockages are experienced differently in various cells, they are likely to share the same characteristics, e.g., to have similar duration and attenuation. Capturing these characteristics using a shared inductive bias, allows us to accelerate the training of predictors for new cells.

In order to account for the survival time or other latency constraints, we consider the problem of predicting whether a link will be blocked within a given time window. Two types of events are predicted: (i) occurrence of at least one blockage within the time window, and (ii) blockage through the entire time window. These predictors are motivated by different applications, as (i) is suitable for applications that are intolerant to any packet errors and (ii) is appropriate if the application is sensitive to consecutive packet losses.

\section{System Model and Problem Definition}\label{sec:sysmodel}
We consider a wireless system comprising a generic manufacturing cell with $K$ spatially co-located devices, see \cref{fig:introfig}. Within the cell, communication is organized in periodic cycles that involve transmissions between each of the $K$ devices and a base station (BS) through a high-rate mmWave link. The link model uses block fading with memory introduced by the physical environment through blockages~\cite{rappaport17}.

The cell environment is described by a random variable $\pi$, which is assumed to be drawn independently from an unknown distribution $\pi\sim p(\pi)$. For example, in the evaluation scenario described in \cref{sec:blockagegen} and depicted in \cref{fig:blockagescenario_pic}, we will use $\pi$ to describe parameters such as the location of the devices and the number and speed of blockage objects.

In the cell characterized by variables $\pi$, the signals received by the BS from device $k$, or vice versa, in cycle $t$ are
\begin{equation}
    \mathbf{y}_{k,t} = h_{k,t}\mathbf{x}_{k,t}+\mathbf{w}_{k,t},
\end{equation}
where $\mathbf{x}_{k,t}$ is the packet transmitted with normalized power $E[\|\mathbf{x}_{k,t}\|_2^2]=1$ (the dimension is not relevant for our discussion), $\mathbf{w}_{k,t}\sim\mathcal{CN}(\mathbf{0},\mathbf{I})$ is additive white Gaussian noise, and $h_{k,t}$ is the scalar fading coefficient. The fading coefficients $h_{k,t}$ are assumed to be realizations from a random process $h_{k,1:t}=h_{k,1},h_{k,2},\ldots,h_{k,t}$ that depends on $\pi$. The channel processes are generally statistically dependent across time steps and devices in the cell. Specifically, the $K$ fading processes $\mathbf{h}_{1:t}=(h_{1,1:t},\ldots, h_{K,1:t})$ are characterized by an unknown joint distribution $p(\mathbf{h}_{1:t}|\pi)$.

We assume that the application requirements of the communication link can be satisfied as long as the instantaneous SNR, $\gamma_{k,t} = |h_{k,t}|^2$, is above a certain threshold $\gamma_0$, which is assumed to be the same for all devices. Conversely, communication fails when $\gamma_{k,t}\le \gamma_0$.

Our goal is to predict future blockages between the BS and a device indexed by $k$ based on previous observations of the channel and, possibly, on side information. The sequence of observations is denoted $\mathbf{o}_{1:t}$, where each observation vector $\mathbf{o}_{t}$ may contain, among others, SNR estimates, device locations, and images captured by cameras~\cite{alrabeiah20,charan21}, at time steps $1,2,\ldots,t$. All observations are available at the BS, which runs the predictive models and takes actions upon a predicted blockage.
While our method is general, we showcase the potential of the approach by predicting at time $t$ the probability that either \emph{any} or \emph{all} of the slots in a time window $[t+\xi,t+\xi+\tau]$ for some time lag $\xi\ge 0$ are blocked. We refer to $\tau$ as the \emph{prediction interval} and $\xi$ as the \emph{prediction delay}. The values $\tau$ and $\xi$ can emerge from timing constraints within the system: For example, if a blockage is predicted, then a controller can switch to a more robust control strategy or a semi-autonomous operation. We shall denote the collection of prediction statistics for the $K$ devices by $\mathbf{z}_t=(z_{1,t},\ldots,z_{K,t})$.

The ideal \emph{any}-predictor is given by the posterior probability
\begin{align}
  f_k^{\cup}(\mathbf{o}_{1:t})&=
  \begin{multlined}[t]
    \Pr\big(\{\gamma_{k,t+\xi+1}\le \gamma_0\} \cup \cdots\\
    \cup \{\gamma_{k,t+\xi+\tau}\le \gamma_0\} \mid \mathbf{o}_{1:t},\pi\big)
  \end{multlined}\\
&=\Pr\left(z_{k,t}^{\cup}=1 \mid \mathbf{o}_{1:t},\pi\right),\label{eq:any_pred}
\end{align}
for $k=1,\ldots,K$ where, for brevity, we introduced the outage variable $z_{k,t}^{\cup}=\mathbbm{1}[\{\gamma_{k,t+\xi+1}\le \gamma_0\} \cup \cdots \cup \{\gamma_{k,t+\xi+\tau}\le \gamma_0\}]$.

The ideal \emph{all}-predictor is similar, but with the intersection of events instead of the union
\begin{align}
  f_k^{\cap}(\mathbf{o}_{1:t})&=
  \begin{multlined}[t]
    \Pr\big(\{\gamma_{k,t+\xi+1}\le \gamma_0\} \cap \cdots\\
    \cap \{\gamma_{k,t+\xi+\tau}\le \gamma_0\} \mid \mathbf{o}_{1:t},\pi\big)
  \end{multlined}\\
  &=\Pr\left(z_{k,t}^{\cap}=1 \mid \mathbf{o}_{1:t},\pi\right),\label{eq:all_pred}
\end{align}
where the outage variable $z_{k,t}^{\cap}$ is similarly defined. Compared to the \emph{any}-predictor, the \emph{all}-predictor is related to the survival time of the system and is useful for systems that are sensitive to consecutive errors. To simplify the notation, we will omit the union and intersection symbols from the notation when the results are valid for both cases.

The observations $\mathbf{o}_{1:t}$ and the prediction statistics $\mathbf{z}_{1:t}$ are random variables drawn from an unknown conditional distribution $p(\mathbf{z}_{1:t}, \mathbf{o}_{1:t} | \pi)$. To enable learning of the unknown predictors \eqref{eq:any_pred} and \eqref{eq:all_pred}, we choose as part of the inductive bias a model class of parametric functions, such as neural networks, and we assume the availability of a dataset of historical observations and target variables from several factory cells (i.e., realizations of $\pi$).
Furthermore, to exploit the dependency between different cells, as defined by the unknown distribution $p(\pi)$, we frame the setting as a \emph{meta-learning} problem.
To be aligned with the meta-learning literature, we refer to the set of historical observations from a single cell as a \emph{task}.

We assume that a \emph{meta-training dataset} $\mathcal{D}=\{\mathcal{D}_1,\ldots,\mathcal{D}_N\}$ is available comprising sequences of length $T_{\text{train}}$ from $N$ meta-training tasks $\mathcal{D}_n=(\mathbf{o}_{1:T_{\text{train}}}^{(n)}, \mathbf{z}_{1:T_{\text{train}}}^{(n)})$. Each meta-training task corresponds to a different factory cell. The meta-training dataset is assumed to be available offline to optimize a training procedure that enables effective refinement based on short training sequences. Following the meta-learning literature, we refer to the new task as \emph{meta-test} task, and denote the training set of a meta-test task $n'$ by $\mathcal{D}_{n'}=(\mathbf{o}_{1:T_{\text{test}}}^{(n')}, \mathbf{z}_{1:T_{\text{test}}}^{(n')})$.

\section{Meta-Learning Predictor}\label{sec:metalearning}
To capture the memory in the fading process, we approximate the ideal posterior distribution $f_k(\mathbf{o}_{1:t})$ in \cref{eq:any_pred,eq:all_pred} via RNNs. We specifically consider RNNs composed of a number of input layers followed by recurrent layers and a number output layers. In general, the parameterized RNN based predictive functions can be written as
\begin{align}
    f_{\varphi_k}(\mathbf{o}_{1:t}) &= \sigma\left(g_{\varphi_k}^{(\text{out})}\left(g_{\varphi_k}^{(\text{rec})}\left(g_{\varphi_k}^{(\text{in})}\left(\mathbf{o}_{t}\right), \mathbf{s}_{k,t-1}\right)\right)\right),\\
    \mathbf{s}_{k,t}&=\eta_{\varphi_k}\left(g_{\varphi_k}^{(\text{in})}\left(\mathbf{o}_{t}\right), \mathbf{s}_{k,t-1}\right),
\end{align}
where $g_{\varphi_k}^{(\text{in})}(\cdot),g_{\varphi_k}^{(\text{rec})}(\cdot),g_{\varphi_k}^{(\text{out})}(\cdot)$ are the the input, recurrent and output layers, respectively; $\sigma(x)=1/(1+e^{-x})$ is the sigmoid function; and $\mathbf{s}_{k,t}\in\mathbb{R}^d$ represents the internal state of the recurrent layer. Common RNN
models the Long Short-Term Memory (LSTM) and the Gated Recurrent Unit (GRU)~\cite{goodfellow2016deep}.

As a measure of prediction accuracy, we consider the \emph{weighted binary cross entropy} (BCE) loss function
\begin{equation}
  \ell(z, x)=-w z \log(x) - (1-z)\log(1-x),\label{eq:bceloss}
\end{equation}
where positive examples are multiplied by the constant $w>0$ to compensate for a potentially imbalanced number of positives and negatives~\cite{cui2019class}. The population loss for a given task characterized by variables $\pi$ is then
\begin{equation}
  L(\mathbf{\varphi})=E_{p(\mathbf{z}_{1:t},\mathbf{o}_{1:t}|\pi)}\left[\frac{1}{K}\sum_{k=1}^K \ell(z_{k,t}, f_{\varphi_k}(\mathbf{o}_{1:t}))\right],
\end{equation}
where the expectation is over the unknown distribution $p(\mathbf{z}_{1:t},\mathbf{o}_{1:t}|\pi)$ and $\mathbf{\varphi}=(\varphi_1,\ldots,\varphi_K)$ are the model parameters.

The population loss $L(\mathbf{\varphi})$ is unknown and is in practice replaced by an empirical estimate obtained from data. Because the function parameters $\varphi_k$ are learned per-device, let us define $\tilde{\mathcal{D}}=\{\tilde{\mathcal{D}}_{1,1},\ldots,\tilde{\mathcal{D}}_{1,K},\tilde{\mathcal{D}}_{2,1},\ldots,\tilde{\mathcal{D}}_{N,K}\}$ as the dataset of observation and target sequences used for each individual device, i.e., $\tilde{\mathcal{D}}_{n,k}=(\mathbf{o}_{1:T_{\text{train}}}^{(n)}, z_{k,1:T_{\text{train}}}^{(n)})$ comprises all observations in task $n$ (e.g., SNRs of all devices) but the target variable only for device $k$. The empirical loss function for device $k$ in task $n$ can then be written
\begin{equation}
  \mathcal{L}^{(t)}_{\mathcal{D}_{n,k}}(\varphi) = \sum_{i=1}^{t} L(z_{k,i}^{(n)},f_{\varphi}(\mathbf{o}_{1:i}^{(n)})).\label{eq:empiricalloss}
\end{equation}
  
The quality of a model trained by minimizing the training loss \eqref{eq:empiricalloss} depends on the number of data points, $t$, observed so far. In order to reduce the data requirements and improve the prediction at earlier time steps $t$, we apply Model-Agnostic Meta-Learning (MAML)~\cite{finn17}. MAML finds a set of model parameters $\theta$, from which the device-specific model parameters $\varphi$ can be obtained by taking one or more gradient steps computed using the $t$ available training samples in $\mathcal{D}_{n',k}$
\begin{equation}
    \mathbf{\varphi} \gets \theta + \beta \nabla_{\theta}\mathcal{L}^{(t)}_{\mathcal{D}_{n,k}}(\theta),
\end{equation}
where $\beta$ is the learning rate. The initialization $\theta$ is optimized via stochastic gradient descent as illustrated in \cref{alg:maml}.

\begin{algorithm}
  \caption{Model-Agnostic Meta-Learning (MAML).}\label{alg:maml}
\begin{algorithmic}[1]
  \algrenewcommand\algorithmicwhile{\textbf{While}}
  \algrenewcommand\algorithmicforall{\textbf{For each}}
  \State \textbf{Input:} Datasets $\tilde{\mathcal{D}}$, meta batch size $B$, step sizes $\alpha, \beta$.
  \State Randomly initialize $\theta$
  \While{convergence criterion not met}
  \State Draw meta-batch $\tilde{\mathcal{D}}'$ of $B$ datasets from $\tilde{\mathcal{D}}$
  \ForAll{dataset $\tilde{\mathcal{D}}_{n,k}$ in $\tilde{\mathcal{D}}'$}
  \State Split $\tilde{\mathcal{D}}_{n,k}$ into two datasets $\tilde{\mathcal{D}}_{n,k}^{\mathrm{tr}}$ and $\tilde{\mathcal{D}}_{n,k}^{\mathrm{te}}$
  \State Compute $\varphi_{n,k} \gets \theta - \alpha\nabla_{\theta}\bar{\mathcal{L}}_{\tilde{\mathcal{D}}_{n,k}^{\mathrm{tr}}}(\theta)$\label{maml_grad1}
  \EndFor
  \State Compute $\theta \gets \theta - \beta\nabla_{\theta}\sum_{n,k\in\tilde{\mathcal{D}}'}\mathcal{L}_{\tilde{\mathcal{D}}_{n,k}^{\mathrm{te}}}(\varphi_{n,k})$\label{maml_grad2}
  \EndWhile
\end{algorithmic}
\end{algorithm}

\section{Evaluation Methodology}\label{sec:evalmethod}
\subsection{Blockage Generation}\label{sec:blockagegen}
We model the channel coefficients as Rician fading channels with magnitudes $|h_{k,t}|$ distributed according to
\begin{equation}
    p_{|h_{k,t}|}(x) = \frac{x}{\sigma_{k}^2}\exp\left(-\frac{x^2+A_{k,t}^2}{2\sigma_k^2}\right)I_0\left(\frac{A_{k,t} x}{\sigma_k^2}\right),
\end{equation}
where $I_0(\cdot)$ is the modified Bessel function of the first kind and order zero, $A_{k,t}$ is the signal amplitude of the dominant path and $\sigma_k$ is the standard deviation of the resultant amplitude from the remaining paths. To incorporate blockage effects, we express the dominant path amplitude as $A_{k,t}=\zeta_{k,t}\bar{A}_k$, where $\zeta_{k,t}$ models the attenuation caused by blockages and $\bar{A}_k$ is the average SNR of device $k$ when the signal is not blocked. The expected SNR at time $t$ is then
\begin{equation}
    \gamma_{k,t}=\zeta_{k,t}^2\bar{A}_k^2+2\sigma_k^2.
\end{equation}

\begin{figure}
  \centering
  \begin{subfigure}[b]{0.9\linewidth}
    \centering
    \includegraphics{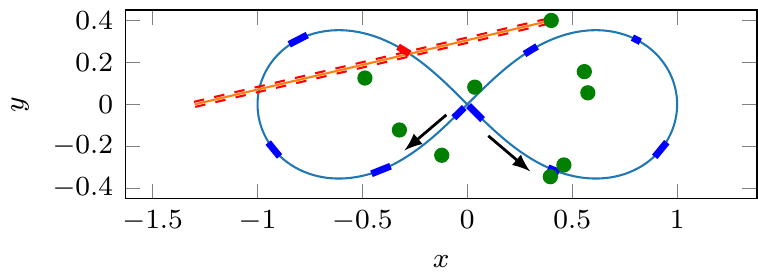}\vspace{-1em}
    \caption{}
    \label{fig:blockagescenario_pic}
  \end{subfigure}
  \begin{subfigure}[b]{0.9\linewidth}
    \centering
    \includegraphics{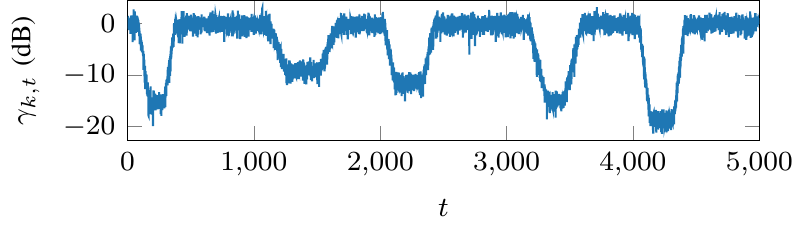}\vspace{-2em}
    \caption{}
    \label{fig:blockagescenario_snr}
  \end{subfigure}    
  \caption{\subref{fig:blockagescenario_pic} Illustration of the blockage generation process with $K=10$ devices (dots), $M=10$ blockage objects (bold segments on the lemniscate), and BS at $(-1.3, 0.0)$. The beamwidth is indicated by the dashed lines. \subref{fig:blockagescenario_snr} The SNR sequence generated for the device at $(0.4,0.4)$.}
  \label{fig:blockagescenario}
\end{figure}

As seen in \cref{fig:blockagescenario}, we assume the $K$ devices are static and uniformly distributed within a rectangular area $[-1,1]\times [-0.5,0.5]$, and that the BS is located at $(-1.3,0)$. The line-of-sight path is assumed to be dominant and is modelled as a rectangular pencil beam with beamwidth $\vartheta$. This assumption is reasonable when the objects are concentrated in a small area at some distance from the BS, where the beamwidth is approximately constant. The blockage attenuation coefficients $\zeta_{k,t}$ are generated as follows. A (random) number $M$ of blockage objects are moving on the slope of an $\infty$-shaped Bernoulli lemniscate centred at the origin and with unit half-width. Each blockage object has a random length (but zero width) and moves with a random (but constant) speed. When one of the blockage objects intersects with a line-of-sight path, the signal amplitude of the dominant path is reduced depending on the attenuation factor of the blockage object. Specifically, we adopt a linear diffraction model inspired by measurements in the mmWave spectrum~\cite{rappaport17}, and model the attenuation coefficients $\zeta_{k,t}$ as the percentage of the beam that is blocked, i.e.,
\begin{equation}
  \zeta_{k,t} = \prod_{m=1}^M \left(\delta_m\right)^{p_{k,m,t}},
\end{equation}
where $\delta_m$ is the attenuation factor of object $m$ and $p_{k,m,t}$ is the percentage of the beam cross section that the object blocks at time $t$. Note that we model the attenuation as a function of the blockage region, and not of the angle between the receiver and the blockage object as in~\cite{rappaport17}. For simplicity, we ignore the fact that multiple objects may block the same part of the beam, and simply aggregate the contributions from the individual blockage objects. Furthermore, we neglect effects such as reflections. The location of the devices, the number of blockage objects and their lengths and speeds and attenuation factors define the random variable $\pi$. Although the setting is clearly simplistic, its main purpose is to model the correlation between device SNRs across time and devices. A snapshot of a sample environment with $K=10$ devices and $M=10$ blockage objects is illustrated in \cref{fig:blockagescenario_pic}, and \cref{fig:blockagescenario_snr} shows the SNR sequence generated for the device at $(0.4,0.4)$. Note the impact of the different object lengths and attenuation factors on the SNR during blockages.

\subsection{Observation Model}
While the proposed algorithm can be used with any observation vector, we evaluate its performance using only SNR observations from all $K$ devices within a factory cell. These are available in most existing wireless systems, and capture the temporal and spatial correlation among blockages. We assume that the SNR of a device can only be obtained if its SNR is higher than $\gamma_0$, and indicate the availability of an SNR measurement by a binary feature in the observation vector. Thus, the observations at each time step are represented as the $2K$-dimensional vector $\mathbf{o}_{t}\in\mathbb{R}^{2K}$ obtained by concatenating the $K$ $(\mathbbm{1}[\gamma_{k,t}\le 0], \gamma_{k,t})$-tuples for $k=1,\ldots,K$. To impose structure that can be beneficial in meta-learning, we put the tuple of the device that we aim to predict at the top of the observation vector. The order of the remaining $K-1$ tuples is arbitrary but fixed.

\subsection{Baseline Models}
We consider three reference baseline models. The first is \emph{na\"ive forecasting}, where the predictor outputs the most recent observed value, that is, $f_k^{\text{naive}}(\mathbf{o}_{k,1:t})=\mathbbm{1}[\gamma_{k,t}\le 0]$. The second baseline is \emph{joint learning}, whereby the RNN is trained on the meta-training dataset in a task-agnostic manner (i.e., without MAML) as in typical machine learning applications. Finally, we also consider \emph{random initialization}, where the network weights are randomly initialized prior to the evaluation, i.e., without using the meta-training dataset.

\section{Results}\label{sec:results}
We evaluate the model on tasks with either $K=20$ or $K=50$ devices generated with the parameters listed in \cref{tab:evalparams}. The specific carrier frequency is irrelevant as it is abstracted in the other parameters. The predictors $f_{\varphi_k}(\mathbf{o}_{1:t})$ are RNNs comprised of a fully connected input layer $g_{\varphi_k}^{(\text{in})}(\cdot)$ with 128 rectified linear units (ReLUs); $g_{\varphi_k}^{(\text{rec})}(\cdot)$ and $\eta_{\varphi_k}(\cdot)$ are given by an LSTM layer with 128 hidden units (see e.g., \cite[Eqs.~(10.40)--(10.44)]{goodfellow2016deep} for details); and finally $g_{\varphi_k}^{(\text{out})}$ is a connected layer with 128 rectified linear units followed by a linear output layer with a single sigmoid-activated output. To compensate for the fact that only around $3\%$ of the target values in the dataset are positive, we weight the positive examples by $w=9$ in \cref{eq:bceloss} (higher values of $w$ did not yield better results).

We train the model on data from $N=100$ tasks, i.e. realizations of $\pi$, each containing $20$ or $50$ devices so that the per-device dataset, $\tilde{\mathcal{D}}$, contains a total of $2000$ or $5000$ sequences. Each sequence contains $T_{\text{train}}=10000$ samples, which for MAML are evenly split into a meta-training and meta-testing sub-sequence. In both joint learning and MAML, we use truncated backpropagation through time with a truncation length of 128 samples. Furthermore, to reduce the training time of MAML we split the samples from each device, $\tilde{\mathcal{D}}_{n,k}^{\text{tr}}$ and $\tilde{\mathcal{D}}_{n,k}^{\text{te}}$, into batches of 512 samples for which the loss can be computed in parallel (in lines \ref{maml_grad1} and \ref{maml_grad2} of \cref{alg:maml}). The consequence is that the LSTM state is not maintained for more than $512$ samples, thus it is only done during training and not in validation and testing.
We use a separate dataset of the same size for testing, and use sequences of length $T_{\text{test}}=10000$ for model adaptation and $10000$ for evaluation.
We note that the amount of training data generally influences the performance of the predictors, as the inductive bias learned by MAML can lead to potential degradation when sufficient data are available~\cite{simeone20}.

\begin{table}
\caption{\textsc{Evaluation Parameters}}
\label{tab:evalparams}
\centering
\begin{tabularx}{0.9\linewidth}{>{\hsize=1.3\hsize}X>{\hsize=0.7\hsize}X}
\toprule
Parameter     & Value\\
\midrule
Average unblocked SNR, $\bar{A}_k^2+2\sigma_k^2$ & 0 dB  \\ 
$K$-factor, $\bar{A}_k^2/(2\sigma_k^2)$ & 15 dB \\
Beamwidth, $\vartheta$ & $0.025$ \\
Blockage SNR threshold, $\gamma_0$ & -20 dB \\
Number of blockage objects & $\mathcal{U}\{2,5\}$   \\
Lengths of blockage objects & $0.05$ \\
Speed of blockage objects (loops/sec.) & $\mathcal{U}(0.005, 0.01)$   \\
Attenuation factors, $\delta_m$ & $\mathcal{U}(-30, -10)$ dB   \\ 
Transmission interval (ms) & 50 \\
\bottomrule
\end{tabularx}
\end{table}

\begin{figure}
  \centering
  \includegraphics{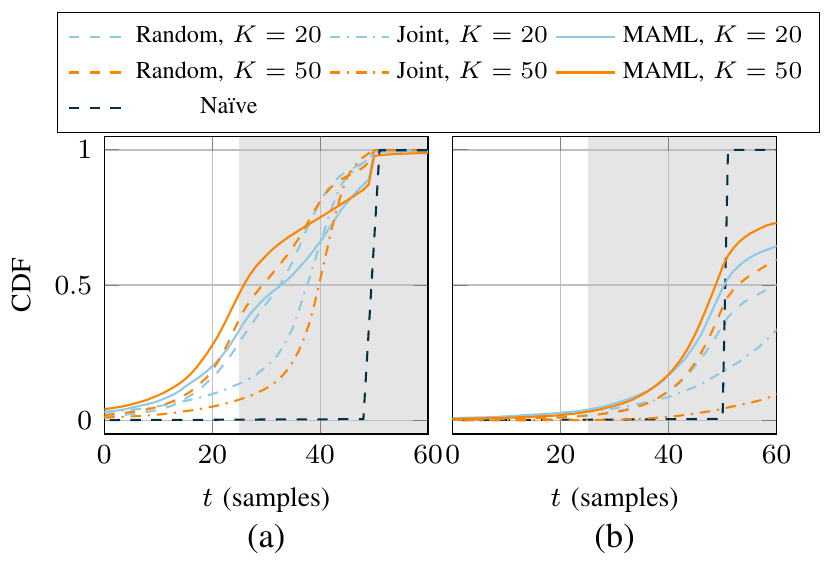}
  \caption{CDF of prediction time for the (a) \emph{any}-predictor ($\xi=0$, $\tau=25$) and (b) \emph{all}-predictor ($\xi=25$, $\tau=3$) with $N=1000$. The detection threshold is $f_k(\mathbf{o}_{k,1:t}) > 0.5$ and the blockage starts at $t=50$ so the predictors should output the blockage from $t=25$.}
  \label{fig:cdf_pred}
\end{figure}

\cref{fig:cdf_pred}(a) shows the CDF of the prediction time for the \emph{any}-predictor with a prediction interval of $\tau=25$ and prediction delay $\xi=0$, where a blockage occurs at time $t=50$. Thus, the predictor is expected to predict the blockage from $t=25$. As can be seen, the MAML predictor generally predicts the blockage earlier than the predictors with randomly and jointly learned initialization, indicating that the MAML procedure learns an inductive bias that is beneficial for efficient learning. The randomly initialized predictor performs better than the jointly learned initialization for both $K=20$ and $K=50$, but not as good as MAML. The joint predictor performs the worst of the neural network based models, which indicates that it is unlikely to find a universal criterion that predicts a blockage. Contrary to the MAML and random predictors, the joint predictor performs worse for $K=50$ than for $K=20$.

Under the same conditions, the prediction time CDF for the \emph{all}-predictor with $\tau=3$ and $\xi=25$ is shown in \cref{fig:cdf_pred}(b). Compared to the \emph{any}-predictor, the predictors are generally slower at predicting the blockage, which is likely due to the fact that the overall probability of experiencing $\tau=3$ consecutive blockages is less than the probability of experiencing any blockage in a window of 25 slots, as was the case with the \emph{any}-predictor. Nevertheless, the MAML predictor again generally predicts the blockage earlier than the random and joint predictors.

\begin{figure}
  \centering
  \includegraphics{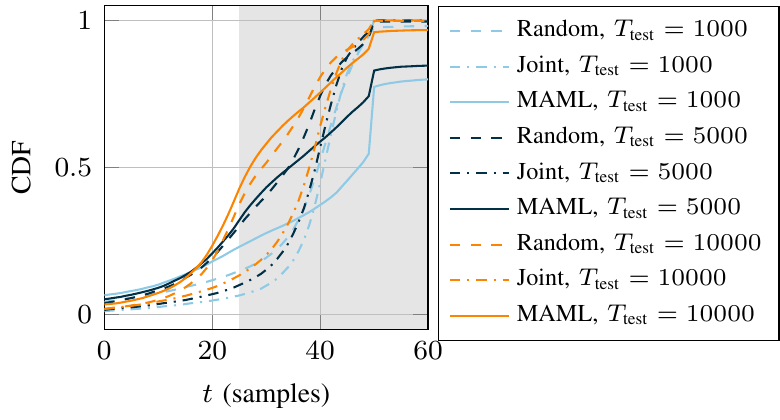}
  \caption{CDF of prediction time for the \emph{any}-predictor for a range of adaptation sequence lengths $T_{\text{test}}$ with $K=50$, $\xi=0$, and $\tau=25$.}
  \label{fig:cdf_pred_vs_ttest}
\end{figure}

\cref{fig:cdf_pred_vs_ttest} shows the impact of the length of the adaptation sequence $T_{\text{test}}$ for the \emph{any}-predictor. MAML outperforms random initialization especially when $T_{\text{test}}$ is small, while the latter approach may be advantageous when $T_{\text{test}}$ is large. In contrast, the jointly trained initialization does not adapt well to the new task. This illustrates the key property that joint learning aims to minimize the average loss across all tasks, whereas MAML tries to minimize the loss of each individual task.

\section{Conclusions}\label{sec:conclusion}
In this paper, we have studied the use of meta-learning to train a recurrent neural network as a blockage predictor for mmWave and THz systems using few samples. While our method is general, we trained the predictors to exploit correlation in SNRs from multiple devices, and considered both an \emph{any}-predictor, which predicts whether there will be at least one blocked transmission within a time window, and an \emph{all}-predictor, which predicts whether the entire window will be blocked. We have shown that meta-learning is beneficial as compared to standard training methods in that it leads to predictors that predict blockages earlier, based on fewer observed data samples.


\end{document}